\pdfoutput=1
\documentclass[reprint, amsmath, amssymb, aps]{revtex4-2}
\usepackage{graphicx}
\usepackage{float}
\usepackage{dcolumn}
\usepackage{color}
\usepackage{latexsym,bm}
\usepackage[normalem]{ulem}
\usepackage{multirow}
\usepackage{appendix}
\usepackage{amsmath}
\usepackage{amsfonts}
\usepackage{booktabs}
\usepackage{array}
\usepackage{soul}
\usepackage{CJKutf8}

\newcommand{\ccc}[1]{\textcolor{black}{{#1}}}
\newcommand{\cc}[1]{\textcolor{black}{{#1}}}

\newcommand{\bra}[1]{\langle #1 |}
\newcommand{\ket}[1]{| #1 \rangle}

\begin{document}
\hbadness=10000
\hyphenpenalty=5000
\tolerance=1000
\begin{CJK}{UTF8}{gkai} 

\title{\cc{Effects of Non-local Pseudopotentials on the Electrical and Thermal Transport Properties of Aluminum: A Density Functional Theory Study}}

\author{Qianrui Liu (刘千锐)}
\affiliation{HEDPS, CAPT, College of Engineering and School of Physics, Peking University, Beijing 100871, P. R. China}
\author{Mohan Chen (陈默涵)}
\email{mohanchen@pku.edu.cn (Corresponding author)}
\affiliation{HEDPS, CAPT, College of Engineering and School of Physics, Peking University, Beijing 100871, P. R. China}
\affiliation{AI for Science Institute, Beijing 100080, P. R. China}
\date{\today}

\begin{abstract}
{\cc{
Accurate prediction of electron transport coefficients is crucial for understanding warm dense matter. Utilizing the density functional theory (DFT) with the Kubo-Greenwood formula is widely used to evaluate the electrical and thermal conductivities of electrons.
\ccc{
By adding the non-local potential correction term that appears in the dynamic Onsager coefficient and using two different norm-conserving pseudopotentials, we predict the electrical and thermal conductivities of electrons for liquid Al (1000 K) and warm dense Al (0.2 to 10 eV).
We systematically investigate the effects of non-local terms in the pseudopotentials and the frozen-core approximation on the conductivities.
We find that taking into account the non-local potential correction and validating the frozen core approximation is essential for accurately calculating the electrical and thermal transport properties of electrons across a wide range of temperatures. }
}
}

\end{abstract}

\maketitle
\end{CJK}

\section{Introduction}
\cc{
Warm dense matter (WDM) is a state of matter characterized by extremely high temperature, high density, and high pressure that commonly exists and plays a significant role in the interiors of giant planets~\cite{99S-Guillot,80AJSS-Wesemael} and inertial confinement fusion~\cite{95PP-Lindl}.
Consequently, research on WDM has emerged as one of the most prominent areas in high energy density physics. Nonetheless, generating and analyzing WDM in laboratory settings presents a significant challenge, making computational modeling an indispensable tool for understanding this unique state of matter. Furthermore, given that WDM comprises partially ionized electrons and strongly coupled ions, incorporating quantum mechanics is essential for accurately simulating and investigating its properties.
Nevertheless, even with the use of modern high-performance computing, it is still challenging to simulate the structural, dynamical, and transport properties of WDM at extremely high temperatures~\cite{20PP-Bonitz,23-White}.
}
%

\cc{
In recent years, various first-principles methods, such as the Kohn-Sham density functional theory (KSDFT)~\cite{64PR-Hohenberg,65PR-Kohn,65PR-Mermin}, the orbital-free density functional theory (OFDFT)~\cite{27cps-Thomas,27ANL-Fermi}, and the path integral Monte Carlo~\cite{84B-Pollock,86L-Ceperley,13L-Brown}, have been developed to investigate WDM across a wide range of temperatures and pressures~\cite{14E-Hu,15JCP-Driver,18L-Ding,18L-Mo,20B-Mo,21E-Militzer,12B-Vlcek,15E-Sjostrom,18PP-Witte,20JPCM-Liu,21MRE-Liu}. Additionally, several KSDFT-based methods have been proposed to overcome the challenge of simulating WDM, including the extended first-principles molecular dynamics~\cite{16PP-Zhang,16B-Gao,20PP-Blanchet,21B-Liu}, the stochastic DFT~\cite{13L-Baer,18B-Cytter}, and the mixed deterministic-stochastic DFT~\cite{20L-White,22B-Liu,24MRE-Chen}, etc.
Within the framework of DFT, the exchange-correlation (XC) functionals considering temperature effects have been developed~\cite{14L-Karasiev,18L-Karasiev} to yield more accurate results for the matter under extreme conditions.
In addition, machine-learning-based molecular dynamics~\cite{18L-Zhang,18ANIPS-Zhang}, known for their ability to efficiently simulate large systems while retaining high accuracy, have recently been widely utilized to investigate WDM with larger systems or longer trajectories~\cite{21R-Zeng,22B-Schorner,22B-Schorner2,20JPCM-Liu,20PP-Yuzhi,21MRE-Liu,24MRE-Chen}.
}

\cc{Several methods have been developed to describe the ion-electron interactions within the framework of DFT. First, adopting norm-conserving pseudopotentials (NCPPs)~\cite{79L-Hamann} in DFT calculations provides an accurate yet efficient way to describe the behavior of valence electrons while neglecting the fast-varying, highly localized core electron wavefunctions. 
Among the NCPPs, the Kleinman-Bylander (KB) approach~\cite{82L-Kleinman} employs a single local radial potential and a few separable $l$-dependent non-local projectors.
Recently, Hamann~\cite{13B-Hamann,17B-Hamann} has developed optimized norm-conserving Vanderbilt (ONCV) pseudopotentials with two projectors, which enhances the accuracy and lower cutoff energies when compared to the traditional NCPPs~\cite{15CPC-Schlipf,18CPC-Setten}.
Second, the ultrasoft pseudopotentials (USPPs)~\cite{90B-Vanderbilt} remove the norm-conserving condition and provide a smoother potential to describe the electron-ion interaction, allowing for a lower energy cutoff and thus reducing computational costs.
Third, another notable approach is the projector augmented-wave (PAW) method~\cite{94B-Blochl}, 
which allows recovery of all-electron quantities while still benefiting from reduced computational costs compared to the all-electron methods.
Although the USPPs and PAW methods can reduce the energy cutoff and improve efficiency, they come with additional complexity in the implementation of certain calculations, such as density-functional perturbation theory (DFPT)~\cite{01RMP-Baroni}.
On the other hand, NCPPs remain widely used due to their simplicity and ease of implementation. In this work, we adopt the NCPPs.
}

\cc{
The electrical and thermal transport properties are essential in fields such as laser heating~\cite{03B-Ivanov}, hydrodynamic instability~\cite{98PP-Marinak}, and metal-nonmetal transitions~\cite{12B-Korobenko}.
Both electrons and ions contribute to the electrical and thermal transport, but the former becomes more important in WDM.~\cite{21MRE-Liu}
In this regard, we focus on calculating the electrical and thermal conductivities driven by electrons. 
To do this, KSDFT-based molecular dynamics and the Kubo-Greenwood (KG) method are employed, which have been successfully applied to various systems, including liquid metals~\cite{05B-Recoules}, silica~\cite{04B-Laudernet}, and plastics~\cite{12E-Lambert}, etc. In particular, various studies have adopted this method to study transport properties of WDM and obtained valuable results~\cite{12B-Vlcek,15E-Sjostrom,18PP-Witte,21MRE-Liu}.
}

\cc{
Notably, when employing the KG formula, the presence of non-local potentials in NCPPs 
leads to a correction term \ccc{in this formula} ({\it vide infra}).
Unfortunately, some previous works have overlooked the influences of these corrections on the computed properties~\cite{04B-Alemany,19PP-Kang,21MRE-Liu}.
To the best of our knowledge, when the NCPPs are employed to study the conductivities of WDM, the effects of non-local potential corrections \ccc{that appear in the Kubo-Greenwood formula} are still inconclusive. In this regard,
\ccc{
we have implemented the non-local potential corrections with the plane wave basis sets and under periodic boundary conditions in the ABACUS package\cite{10JPCM-Mohan,16CMS-Li,22B-Liu}. Next, we select Al at two densities (2.35 and 2.70 g/cm$3$) and the temperature ranges from 0.086 to 10 eV.
We first compute the electrical and thermal conductivities of electrons for aluminum (Al) with the revised formula. 
Next, we thoroughly investigate the effects of using two different pseudopotentials with or without the non-local potential corrections on the computed conductivities. 
Finally, We analyze the density of states, the decomposed electrical conductivity, and the Lorents number of Al across a wide range of temperatures and elucidate our findings.}
}

\cc{
The paper is organized as follows.
In Sec.~II, we provide a brief introduction to the Kubo-Greenwood formula and explain the formulas to implement the non-local potential corrections with the plane wave basis set. 
Additionally, we outline the setup for calculating the electrical and thermal conductivities of Al.
In Sec.~III, we present our results and analysis for Al.
The concluding remarks are shown in Sec.~IV.}

\section{Methods}
\subsection{Electrical conductivity and thermal conductivity}
\cc{
To calculate the electrical and thermal conductivities, we first need to compute the dynamic Onsager coefficients $L_{mn}(\omega)$ ($m,n=1,2$) using the Kubo-Greenwood formula. The equation is as follows:
\begin{equation}\label{eq:KG}
	\begin{aligned}
		&L_{mn}(\omega)=(-1)^{m+n}\frac{2\pi e^2}{3\omega\Omega}\\
		&\times\sum_{ij\alpha\mathbf{k}}W(\mathbf{k})\left(\frac{\epsilon_{i\mathbf{k}}+\epsilon_{j\mathbf{k}}}{2}-\mu\right)^{m+n-2}|
		\langle\Psi_{i\mathbf{k}}|\hat{v}_\alpha|\Psi_{j\mathbf{k}}\rangle|^2\\
		&\times[f(\epsilon_{i\mathbf{k}})-f(\epsilon_{j\mathbf{k}})]\delta(\epsilon_{j\mathbf{k}}-\epsilon_{i\mathbf{k}}-\hbar\omega),
	\end{aligned}
\end{equation}	
where $\omega$ represents the frequency, $e$ is the elementary charge, $\Omega$ denotes the volume of the cell, $\mu$ refers to the chemical potential. $\Psi_{i\mathbf{k}}$ represents the wave function of the $i$-th band with $\mathbf{k}$ being a point in the first Brillouin zone, and the corresponding eigenvalue is $\epsilon_{i\mathbf{k}}$. $\hat{v}_\alpha$ denotes the $\alpha$-th component of the velocity operator $\hat{\mathbf{v}}$.
The Fermi-Dirac distribution function of electrons is $f$,
and $\delta$ is the delta function. In practice, the delta function is approximated by a Gaussian function:
\begin{equation}
	\delta(E) \approx \frac{1}{\sqrt{2\pi}\Delta E}\exp\left(-E^2/(2\Delta E^2)\right),
\end{equation}
where $\Delta E$ determines the width of the Gaussian function.
In this work, we set the full width at half maximum ($\mathrm{FWHM} = 2.3548\Delta E$) to 0.1 eV.
}

\cc{
With the above formulas, the electrical conductivity $\sigma$ and thermal conductivity $\kappa$ of electrons respectively take the form of
\begin{equation}\label{eq:sigma}
	\sigma=\lim_{\omega\to0}L_{11}(\omega),
\end{equation}
\begin{equation}\label{eq:kappa}
	\kappa=\lim_{\omega\to0} \frac{1}{e^2T}\left(L_{22}(\omega)-\frac{L_{12}^2(\omega)}{L_{11}(\omega)}\right).
\end{equation}
In addition, to allow for extrapolation to zero frequency, we employ specific functions to fit the frequency-dependent lines, which can be found in Supplementary Material. It is worth noting that the choices of different fitting functions do not substantially affect the results when the fitting range is close to the zero frequency $\omega$=0.}

\subsection{Non-local potential correction}
\cc{
The velocity operator~\cite{71APS-Starace} in quantum mechanics, denoted as $\hat{\mathbf{v}}$, is defined as
\begin{equation}
	\hat{\mathbf{v}} = \frac{i}{\hbar}\left[\hat{H}, \hat{\mathbf{r}}\right],
\end{equation}
Here, $\hat{H}$ represents the Hamiltonian operator, and $\hat{\mathbf{r}}$ represents the position operator.
For Hamiltonian that contains the non-local pseudopotential, an additional commutator enters the formalism as~\cite{71APS-Starace,91B-Read}
\begin{equation}\label{eq:velocity}
	\hat{\mathbf{v}} = \frac{\hat{\mathbf{p}}}{m_e}+\frac{i}{\hbar}\left[\hat{V}_\mathrm{NL},\hat{\mathbf{r}}\right],
\end{equation}
where $\hat{\mathbf{p}}$ is the momentum operator, $m_e$ is the mass of electrons, and $\hat{V}_\mathrm{NL}$ depicts the non-local pseudopotential operator.
When dealing with the Hamiltonian that only contains the local potential terms, the velocity operator can be represented by the momentum operator alone.
However, for systems with non-local potential terms, the correction term in the second part of Eq.~(\ref{eq:velocity}) cannot be ignored.}

In this work, we adopt the plane-wave basis set with the periodic boundary conditions and the $k$-point sampling method~\cite{76B-Monkhorst}, with which the electronic wave function can be written as
\begin{equation}
|\Psi_{i\mathbf{k}}\rangle=\sum_{\mathbf{G}}c_{i\mathbf{k}}(\mathbf{G})|\mathbf{k+G}\rangle,
\end{equation}
where $\mathbf{G}$ depicts the plane wave basis and $\{c_{i\mathbf{k}}\}$ are the coefficients of plane wave basis sets \ccc{for band $i$}. 
We denote the wave vector as $\mathbf{q}=\mathbf{k+G}$, which satisfies $\sum_{\mathbf{q}}|\mathbf{q}\rangle\langle\mathbf{q}|=I$ \ccc{with $I$ being an identity matrix}.
$c_{i\mathbf{k}}(\mathbf{G})$ takes the formula of
\begin{equation}
    c_{i\mathbf{k}}(\mathbf{G})=\langle\mathbf{q}|\Psi_{i\mathbf{k}}\rangle=\Psi_{i\mathbf{k}}(\mathbf{q}).
\end{equation}
\ccc{Thus,} the non-local correction term in the velocity matrix has the form of
\begin{equation}
\begin{aligned}
	&\langle\Psi_{i\mathbf{k}}|[\hat{V}_\mathrm{NL},\hat{\mathbf{r}}]|\Psi_{j\mathbf{k}}\rangle\\	
&=\sum_{\mathbf{qq'}}\langle\Psi_{i\mathbf{k}}|\mathbf{q}\rangle\langle\mathbf{q}|[\hat{V}_\mathrm{NL},\hat{\mathbf{r}}]|\mathbf{q'}\rangle\langle\mathbf{q'}|\Psi_{j\mathbf{k}}\rangle,\\
&=\sum_{\mathbf{qq'}}\Psi_{i\mathbf{k}}^{*}(\mathbf{q})\langle\mathbf{q}|[\hat{V}_\mathrm{NL},\hat{\mathbf{r}}]|\mathbf{q'}\rangle\Psi_{j\mathbf{k}}(\mathbf{q'}).
\end{aligned}
\end{equation}
Since the position operator $\hat{\mathbf{r}}$ is ill-defined in a periodic system, \ccc{the} $[\hat{V}_\mathrm{NL},\hat{\mathbf{r}}]$ \ccc{operator} can be calculated in \ccc{the} reciprocal space~\cite{87B-Hybertsen} as
\begin{equation}
\label{eq:vnlq}
    \langle\mathbf{q}|[\hat{V}_\mathrm{NL},\hat{\mathbf{r}}]|\mathbf{q'}\rangle=(\nabla_{
	\mathbf{q}}+\nabla_{
	\mathbf{q'}})V_{\mathrm{NL}}(\mathbf{q},\mathbf{q'})
\end{equation}

To compute the above term, we need to first calculate the non-local pseudopotential term in the plane wave basis, which is
\begin{equation}
	V_{\mathrm{NL}}(\mathbf{q},\mathbf{q'})=\sum_{\kappa=1}^{N_\mathrm{t}}S^\kappa(\mathbf{q'-q})V_{\mathrm{NL}}^\kappa(\mathbf{q,q'}),
\end{equation} 
where $N_\mathrm{t}$ is the number of atom types, $V_{\mathrm{NL}}^\kappa(\mathbf{q,q'})$  is the non-local pseudopotential of the $\kappa$-th element, and the first part $S^\kappa$ is the structure factor of the $\kappa$-th element, given by
\begin{equation}
S^\kappa(\mathbf{G})=\sum_{j=1}^{N_\kappa}\exp(i\mathbf{G} \cdot \mathbf{\tau}_{\kappa j}),
\end{equation}
where $N_\kappa$ denotes the number of $\kappa$-type atoms and $\mathbf{\tau}$ represents the position of the atom. In this case, we find
\begin{equation}
(\nabla_\mathbf{q}+\nabla_\mathbf{q'})S^\kappa(\mathbf{q'-q})=0.
\end{equation}

\cc{
Furthermore, the second part is
the non-local potential for the $\kappa$-th element and can be calculated using the formula of
\begin{equation}
	V_{\mathrm{NL}}^\kappa(\mathbf{q,q'}) = \sum_{ll'mm'}D_{lm,l'm'}^\kappa\beta^\kappa_{lm}(\mathbf{q})\beta^{\kappa*}_{l'm'}(\mathbf{q'}),
\end{equation}
where $\beta_{lm}^\kappa(\mathbf{q})$ is the non-local projector of NCPPs expanded with the plane wave basis. Additionally, the atom type is $\kappa$ and the coefficient matrix of projectors is $D_{lm,l'm'}$. In detail, $\beta_{lm}^\kappa(\mathbf{q})$ is written as
\begin{equation}
	\beta_{lm}^\kappa(\mathbf{q})=\frac{4\pi(-i)^l}{\sqrt{\Omega}}f_l^\kappa(q)Y_{lm}(\hat{\mathbf{q}}).
\end{equation}
Here, the radial part is obtained by
\begin{equation}
	f_l^\kappa(q) = \int_0^{+\infty} F_l^\kappa(r)j_l(qr)r^2\mathrm{d}r.
\end{equation}
In this equation, $rF_l^\kappa(r)$ is a one-dimensional non-local projector read from the pseudopotential file, $j_l$ depicts the spherical Bessel function and $Y_{lm}(\hat{\mathbf{q}})$ is the spherical harmonic function.}

\cc{
Based on the above formulas, the key part of the non-local potential correction term can be expressed as
\begin{equation}
\label{eq:nablavnl}
\begin{aligned}
	&(\nabla_\mathbf{q}+\nabla_\mathbf{q'})V_{\mathrm{NL}}(\mathbf{q,q'})\\
&=\sum_{\kappa=1}^{N_\mathrm{t}}S^\kappa(\mathbf{q'-q})(\nabla_\mathbf{q}+\nabla_\mathbf{q'})
	V_{\mathrm{NL}}^\kappa(\mathbf{q,q'})\\
&=\sum_{\kappa=1}^{N_\mathrm{t}}S^\kappa(\mathbf{q'-q})\sum_{ll'mm'}D_{lm,l'm'}^\kappa\\
	&\times\Big[\mathbf{g}^\kappa_{lm}(\mathbf{q})\beta^{\kappa*}_{l'm'}(\mathbf{q'})
	+\beta^\kappa_{lm}(\mathbf{q})\mathbf{g}^{\kappa*}_{l'm'}(\mathbf{q'})\Big],
\end{aligned}
\end{equation}
where $\mathbf{g}_{lm}^\kappa(\mathbf{q})$ represents the gradient of $\beta_{lm}^\kappa(\mathbf{q})$ with respect to $\mathbf{q}$ and has the following form
\begin{equation}
	\begin{aligned}
		\mathbf{g}_{lm}^\kappa(\mathbf{q})& = \frac{4\pi(-i)^l}{\sqrt{\Omega}}\\
		&\times\left[q^lY_{lm}(\hat{\mathbf{q}})\nabla_\mathbf{q}\left(\frac{f^\kappa_l(q)}{q^l}\right)
		+\frac{f^\kappa_l(q)}{q^l}\nabla_\mathbf{q}\left(q^lY_{lm}(\hat{\mathbf{q}})\right)\right].
	\end{aligned}
\end{equation}
It should be noted that the gradient of $q^lY_{lm}(\hat{\mathbf{q}})$ is used instead of \ccc{the gradient of} $Y_{lm}(\hat{\mathbf{q}})$ to avoid the singularity at $\mathbf{q=0}$.
}

\cc{
In total, the velocity matrix $\langle\Psi_{i\mathbf{k}}|\hat{v}_\alpha|\Psi_{j\mathbf{k}}\rangle$ in Eq.~(\ref{eq:KG}), can be evaluated in the reciprocal space \ccc{with plane wave basis} as
\begin{equation}
\begin{aligned}
\sum_{\mathbf{qq'}}\Psi_{i\mathbf{k}}^{*}(\mathbf{q})\left(\frac{\hbar\mathbf{q}}{m_e}\delta(\mathbf{q-q'})+\frac{i}{\hbar}\bra{\mathbf{q}} [\hat{V}_{\mathrm{NL}},\mathbf{r}] \ket{\mathbf{q'}}\right)\Psi_{j\mathbf{k}}(\mathbf{q'}),
\end{aligned}
\end{equation}
}
where the second term is calculated through Eqs.~(\ref{eq:vnlq}) and (\ref{eq:nablavnl}).

\subsection{Computational Details}
\cc{
\ccc{With the above formulas implemented,} we performed a systematic study on the electrical and thermal conductivities of Al at temperatures ranging from 0.2 to 10 eV. The densities were chosen to be 2.35 and 2.70 $\mathrm{g/cm^3}$.
First, to obtain atomic configurations of Al at various temperatures, we utilized Born-Oppenheimer molecular dynamics (BOMD) simulations based on OFDFT.
The Wang-Teter (WT) kinetic energy density functional~\cite{92B-Wang} was employed and has been validated in previous works~\cite{20JPCM-Liu,21MRE-Liu}.
During BOMD simulations, the Nos\'{e}-Hoover thermostat~\cite{84JCP-Nose,85A-Hoover} was adopted in the NVT ensemble.
We utilized 256 atoms in a cell for temperatures up to 5 eV. For higher temperatures, we used 108 atoms. The energy cutoff for the electron density was set to 80 Ry.
The simulations were performed for 4000 steps with time steps being $\frac{r_s}{60\bar{v}}$, where $r_s$ was the Weigner-Setz radius and $\bar{v}=\sqrt{T/m}$ was the average velocity of atoms. Here, $T$ is the temperature and $m$ is the mass of an Al atom.
From the BOMD trajectories, we selected 5 atomic configurations to calculate the electrical and thermal conductivities of electrons in the Al system, which has been tested to be converged.}

\cc{
Second, we employed the Kohn-Sham density functional theory (KSDFT) and the Kubo-Greenwood formula, as shown in Eqs.~(\ref{eq:KG}),~(\ref{eq:sigma}), and~(\ref{eq:kappa}), to calculate the Onsager coefficients and subsequently the electrical and thermal conductivities of Al.
We used an energy cutoff of 50 Ry to describe the wave functions in KSDFT calculations.
In particular, we utilized two NCPPs, which are named NC11 (11 valence electrons) and NC3 (3 valence electrons). 
On the one hand,
the NC11 pseudopotential, generated using the optimized norm-conserving Vanderbilt pseudopotential method via the ONCVPSP package~\cite{13B-Hamann,17B-Hamann}, had 11 valence electrons and a cutoff radius of 0.50 \AA. 
On the other hand, the NC3 pseudopotential was generated through the PSlibrary package~\cite{14CMS-Corso} using the Troullier-Martins method~\cite{91B-Troullier}. 
In addition, the NC3 pseudopotential contained 3 valence electrons, and the cutoff radius was set to 1.38 \AA.
To ensure convergence, we selected a larger number of KS bands at higher temperatures, ensuring that the occupation of the last band was smaller than 1e-5.
For temperatures of 1000 K and 0.2 eV, we used a $5\times5\times5$ $k$-point and $3\times3\times3$ $k$-point mesh~\cite{76B-Monkhorst}, respectively.
For temperatures ranging from 0.5 to 1.0 eV, we employed a $2\times2\times2$ $k$-point mesh, and we only considered the $\Gamma$ point at higher temperatures.}

\cc{
Finally, both OFDFT and KSDFT simulations employed the Perdew-Burke-Ernzerhof (PBE)~\cite{96L-PBE} XC functional. A previous work~\cite{21MRE-Liu} had demonstrated that using either PBE or LDA~\cite{65PR-Kohn} XC did not significantly affect the results. Additionally, we also tested a temperature-dependent XC functional called KSDT~\cite{14L-Karasiev,18L-Karasiev} and found that it did not substantially influence the results within the temperature and density range studied in this work (see Supplementary Material).
All of the above simulations were performed using the ABACUS v3.2 package~\cite{abacus,10JPCM-Mohan}.
}

\section{Results and Discussion}
\subsection{Liquid Al}

\begin{figure}
	\centering
	\includegraphics[width=8.6cm]{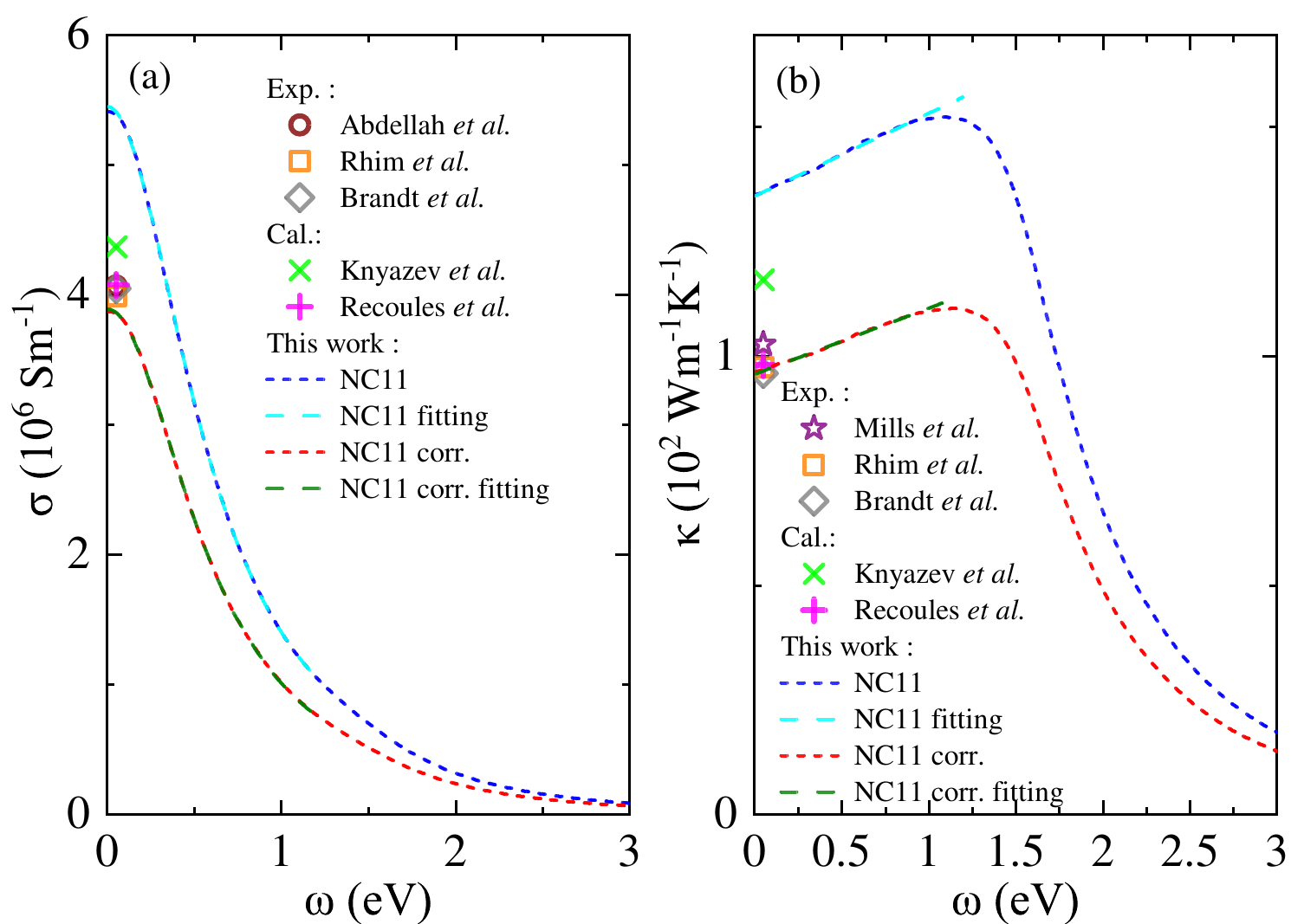}
	\caption{\cc{(a) Frequency-dependent electrical ($\sigma$ in $10^6~\rm{Sm^{-1}}$) and (b) thermal conductivities ($\kappa$ in $\rm{10^2~Wm^{-1}K^{-1}}$) of liquid Al at $T$ = 1000 K. 
 The NC11 label refers to a norm-conserving pseudopotential with 11 valence electrons. The corr. label depicts the line obtained by considering the non-local pseudopotential correction, while the fitting label depicts the line obtained by a fitting method. Here, the electrical conductivities are fitted with a Drude model, while the thermal conductivities are linearly fitted. Computational results from Recoules {\it et al.}~\cite{05B-Recoules} and Knyazev {\it et al.}~\cite{13CMS-Knyazev}, as well as available experimental data at $\omega=0$ eV are shown for comparison~\cite{96IMR-Mills, 98RSI-Rhim,05PM-Abdellah,07IJT-Brandt}.   } }\label{fig:1000K}
\end{figure}

\cc{
Figs.~\ref{fig:1000K}(a) and (b) respectively illustrate the calculated frequency-dependent electrical and thermal conductivities of liquid Al with and without the non-local pseudopotential corrections. 
In addition, first-principles data from previous works ~\cite{05B-Recoules,13CMS-Knyazev} and experimental results are displayed for comparison.
All of the above data are obtained for liquid Al at 1000 K with a density of 2.35 $\mathrm{g/cm^3}$.
Our calculations utilize a simulation cell containing 256 atoms and a 5$\times$5$\times$5 shifted $k$-point mesh to ensure convergence.
We use the NC11 pseudopotential with 11 valence electrons for liquid Al.
In particular, to extrapolate the frequency-dependent electrical and thermal conductivities to zero frequency and obtain the transport properties, we employ the Drude formula to fit the frequency-dependent (dynamic) electrical conductivity and use a linear function to fit the thermal conductivity. The fitting range is from 0.15 to 1.00 eV.
Note that the experimental data of Mills {\it et al.}~\cite{96IMR-Mills} and Rhim {\it et al.}~\cite{98RSI-Rhim} are obtained by substituting 1000 K into the corresponding fitting equation, while the results of Abdellah {\it et al.}~\cite{05PM-Abdellah} and Brandt {\it et al.}~\cite{07IJT-Brandt} are obtained by linearly interpolating their results from the table to 1000 K.
}

\begin{figure*}
	\centering
	\includegraphics[width=18cm]{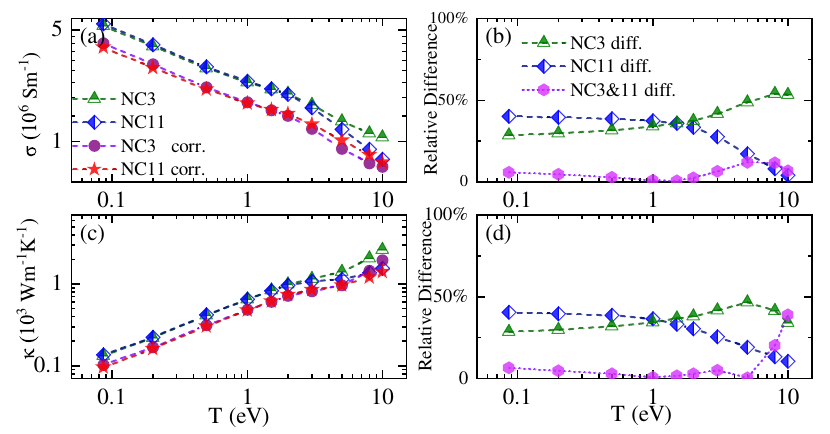}
	\caption{\cc{(a) Electrical ($\sigma$ in $\rm{Sm^{-1}}$) and (c) thermal conductivities ($\kappa$ in $\rm{10^3~Wm^{-1}K^{-1}}$) of liquid and warm dense Al at a density of 2.35 $\mathrm{g/cm^3}$ and temperatures ranging from 0.086 to 10 eV. The simulation cell contains 256 atoms for temperatures no more than 5 eV and 108 atoms for higher temperatures. The NC3 and NC11 labels refer to two norm-conserving pseudopotentials with valence electrons being 3 and 11, respectively. The core. label depicts the data obtained by considering the non-local pseudopotential correction.
 (b) and (d) plot the relative differences of $\sigma$ and $\kappa$, respectively. The label
 ``NCx diff.“ represents the relative difference computed by the formula of $|V_\mathrm{NCx} - V_\mathrm{NCx\ corr.}|/V_\mathrm{NCx\ corr.}$, where $x$ is 3 or 11 and $V$ is $\sigma$ or $\kappa$.
 Additionally, the term ``NC3\&11 diff." depicts $|V_\mathrm{NC3\ corr.} - V_\mathrm{NC11\ corr.}|/V_\mathrm{NC11\ corr.}$.}
} \label{fig:cond35}
\end{figure*}

\cc{
Notably, we find both zero-frequency electrical and thermal conductivities ($\omega=0$) of Al obtained with the non-local potential correction (labeled as NC11 corr. fitting) yield substantially smaller values than the ones without the correction (labeled as NC11 fitting).
In detail, the calculated electrical conductivity at $\omega=0$ with the non-local potential correction shown in Fig.~\ref{fig:1000K}(a) is $3.89\times10^6$ $\mathrm{Sm^{-1}}$, which is 40.1\% lower than the value of $5.45\times10^6$ $\mathrm{Sm^{-1}}$ without correction.
For comparison, the DC electrical conductivity ($\omega=0$) measured by Abdellah {\it et al.}~\cite{05PM-Abdellah}, Rhim {\it et al.}~\cite{98RSI-Rhim}, and Brandt {\it et al.}~\cite{07IJT-Brandt} from experiments are $4.07\times10^6$, $3.99\times10^6$, and $4.05\times10^6$ $\mathrm{Sm^{-1}}$, respectively. 
We conclude that the value with the non-local potential correction matches better with the experimental data.
In Fig.~\ref{fig:1000K}(b), the thermal conductivity at $\omega=0$ calculated with the non-local potential correction (labeled as NC11 corr. fitting) is 96 $\mathrm{Wm^{-1}K^{-1}}$, which is 40.6\% lower than the value of 135 $\mathrm{Wm^{-1}K^{-1}}$ without correction (labeled as NC11 fitting). On the other hand, experimental measurements by Mill {\it et al.}~\cite{96IMR-Mills}, Rhim {\it et al.}~\cite{98RSI-Rhim}, and Brandt {\it et al.}~\cite{07IJT-Brandt} yield values of 103, 98, and 96 $\mathrm{Wm^{-1}K^{-1}}$, respectively.
Again, we find the calculated data with the non-local potential correction agrees well with the experimental results.
}
%


%
\cc{
We also compare the calculated results with previous computational works by Recoules {\it et al.}~\cite{05B-Recoules}, and Knyazev {\it et al.}~\cite{13CMS-Knyazev}. 
In detail, Knyazev {\it et al.}~\cite{13CMS-Knyazev} used 3 valence electrons in a USPP to describe Al, and the non-local potential corrections are not included in the results. More discussions \ccc{are in the later session}. 
On the other hand, Recoules {\it et al.}~\cite{05B-Recoules} utilized \ccc{the density functional perturbation theory (DFPT)} to account for the effects of non-local potential in NCPPs and obtained better results compared to experimental results.
The above results demonstrate the importance of considering the non-local potential correction in computed DC electrical and thermal conductivities for liquid Al.
}



\subsection{Warm dense Al with a density of 2.35 $g/cm^3$}

\cc{
Besides the non-local potential correction, the use of frozen core approximation in pseudopotentials may substantially affect the computed electrical and thermal conductivities of Al. Figs.~\ref{fig:cond35}(a) and \ccc{(c)} show the calculated electrical and thermal conductivities of Al 
in terms of using \ccc{two} different pseudopotentials, respectively.
The temperatures range from 0.086 to 10 eV.
The two used NCPPs are \ccc{named as the NC3 and NC11 pseudopotentials}.
Here, the NC3 and NC11 labels refer to two norm-conserving pseudopotentials with valence electrons being 3 and 11, respectively.
In addition, we compute the electrical and thermal conductivities with and without the non-local pseudopotential corrections. We have the two following findings in Figs.~\ref{fig:cond35}(a) and \ccc{(c)}.
}

Firstly, in the temperature range studied in Fig.~\ref{fig:cond35}(a), the computed electrical conductivity decreases as temperature increases, which can be explained by an increase in the collision frequency of electrons with elevated temperatures. Conversely, Fig.~\ref{fig:cond35}\ccc{(c)} illustrates the computed thermal conductivity becomes larger with rising temperatures. Secondly, comparing the data with and without non-local potential corrections reveals that the corrected values are systematically lower than those without corrections for both electrical and thermal conductivities.

\begin{figure*}
	\centering
	\includegraphics[width=18cm]{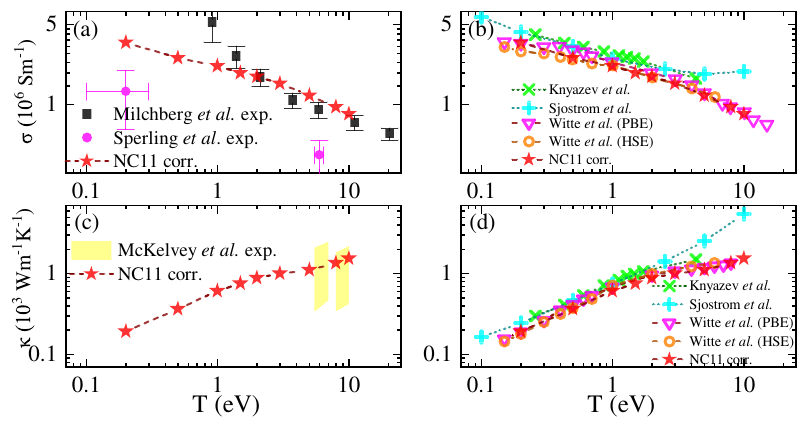}
	\caption{
 \cc{
 (a-b) Electrical ($\sigma$ in $10^6~\rm{Sm^{-1}}$) and (c-d) thermal conductivities ($\kappa$ in $\rm{10^3~Wm^{-1}K^{-1}}$) of liquid and warm dense Al at a density of 2.70 $\mathrm{g/cm^3}$ and temperatures ranging from 0.086 to 10 eV. The simulation cell contains 256 Al atoms for temperatures no more than 5 eV and 108 atoms for higher temperatures. The NC11 labels refer to the norm-conserving pseudopotentials with valence electrons being 11. The “corr." term depicts the calculated data obtained by considering the non-local pseudopotential correction.
 Calculation results from Knyazev {\it et al.}~\cite{13CMS-Knyazev,14PP-Knyazev}, Sjostrom {\it et al.}~\cite{15E-Sjostrom}, and Witte {\it et al.}~\cite{18PP-Witte}, as well as the experimental results from Milchberg {\it et al.}~\cite{88L-Milchberg}, Sperling {\it et al.}~\cite{15L-Sperling}, and McKelvey {\it et al.}~\cite{17SR-McKelvey} are shown for comparison. PBE and HSE refer to two different exchange-correlation functionals used in DFT calculations.
 }
 }\label{fig:cond70}
\end{figure*}

\cc{
Figs.~\ref{fig:cond35} (b) and (d) \ccc{further} illustrate the relative differences for the calculated electrical and thermal conductivities, respectively. 
Interestingly, we find the non-local potential corrections \ccc{for the NC3 pseudopotential} exhibit a trend of becoming substantially larger at around 2 eV and higher temperatures. This can be explained by the fact that the NC3 pseudopotential has 8 fewer electrons than the NC11 pseudopotential \ccc{in the frozen core approximation; the missing 8 electrons in the NC3 pseudopotential affect the computed conductivities at high temperatures,} so the NC3 pseudopotential approximates the ion-electron interactions worse than the NC11 pseudopotential \ccc{at high temperatures. Consequently, the non-local pseudopotential loses transferability and suffers more errors at higher temperatures}. }

\cc{
On the contrary, when the NC11 pseudopotential is used, we observe that the non-local potential corrections tend to exhibit another trend of becoming substantially smaller at around 2 eV and higher temperatures. Since the errors introduced by an insufficient number of electrons are minimized, we attribute this trend to the velocity operator shown in Eq.~\ref{eq:velocity} comprising both the kinetic energy term and non-local potential term of electrons, with the former one becoming increasingly larger than the latter one as the temperature rises.
}

\cc{
We also compare the relative differences of electrical and thermal conductivities obtained by the two different pseudopotentials, i.e., NC3 and NC11. At low temperatures below 2 eV, the two pseudopotentials yield similar results, but as the temperature rises over 2 eV, the differences become increasingly larger, indicating \ccc{again that} the invalidity of the frozen-core approximation for the \ccc{NC3 pseudopotential} at higher temperatures.
}


\subsection{Warm dense Al with a density of 2.7 $g/cm^3$}

\cc{
Figs.~\ref{fig:cond70} (a) and (c) illustrate the 
computed temperature-dependent electrical and thermal conductivities as obtained from the NC11 pseudopotential, respectively. The temperature ranges from \ccc{0.2} to 10 eV while the density of Al is chosen to be $\rho$=2.7 $\mathrm{g/cm^3}$. We also compute the electrical and thermal conductivities by using the NC3 pseudopotential}. In addition, available experimental results are added for comparison. \ccc{Although the trends exhibited by the conductivities at the density of $\rho$=2.7 g/cm$^3$ are similar to those of $\rho$=2.35 g/cm$^3$ shown in Fig.~\ref{fig:cond35}, there are some points worth noting.}

\cc{
We compare our calculated data in Fig.~\ref{fig:cond70} (a) with the experimental data reported by Milchberg {\it et al.}~\cite{88L-Milchberg} and Sperling {\it et al.}~\cite{15L-Sperling}. 
\ccc{We observe that the two experiments exhibit discrepancies in electrical conductivity. In fact,}
accurately measuring the conductivity of WDM from experiments has always been challenging. For example, the experiment of Milchberg {\it et al.} did not directly measure the temperature~\cite{88L-Milchberg} and it was re-estimated by Dharma-wardana {\it et al.}~\cite{92A-Dharma} based on the degree of ionization calculated using KSDFT. In another first-principles work by Mo {\it et al.}~\cite{18L-Mo}, the measured temperature from the experiment of Sperling {\it et al.} is argued to be lower.
%
In general, our calculated electrical conductivities (NC11 corr.) are larger than those from Sperling {\it et al.}~\cite{15L-Sperling}, \ccc{and we suspect the discrepancies may not be caused by the insufficient number of electrons or the lack of non-local potential corrections in the NC11 pseudopotential. The exchange-correlation functional may cause certain errors~\cite{17L-Witte,18PP-Witte}.} Meanwhile, our calculated electrical conductivities align with the data from Milchberg {\it et al.}~\cite{88L-Milchberg} when the temperature is larger than 2 eV. }
However, it is worth noting that Sperling {\it et al.} and Milchberg {\it et al.} are ultrafast experiments, where electrons are heated significantly faster than ions, and some studies suggested that they should be analyzed with a two-temperature model~\cite{92A-Dharma,20E-Wetta,22CPP-Wetta}.

\cc{
Fig.~\ref{fig:cond70}(c) illustrates the thermal conductivities of Al. We compare our calculated results with the experimental data from proton-heated warm dense Al by McKelvey {\it et al.}~\cite{17SR-McKelvey}. 
Our results fall within the error bars of the experimental data for Al densities ranging from 1.7 $\mathrm{g/cm^3}$ to 2.7 $\mathrm{g/cm^3}$.
}

\cc{
Figs.~\ref{fig:cond70} (b) and (d) compare the calculated results with previous first-principles computational works by Knyazev {\it et al.}~\cite{13CMS-Knyazev}, Sjostrom {\it et al.}~\cite{15E-Sjostrom}, and Witte {\it et al.}~\cite{18PP-Witte}.}
\cc{
Our results, incorporating the non-local potential corrections and utilizing NC11 pseudopotential, are systematically lower than the results obtained from Knyazev {\it et al.} To explain the differences,
we notice that a USPP with 3 valence electrons without non-local potential correction was used for Al in the work of Knyazev {\it et al.}
\ccc{For comparison, when using the NC3 and NC11 pseudopotential,} the electrical and thermal conductivities calculated without non-local potential corrections are
systematically higher than those with corrections, as respectively shown in Figs.~\ref{fig:cond35}(a) and (c) for Al density being 
2.35 g/cm$^3$; the conclusion holds when the density increases to 2.7 g/cm$^3$. \ccc{Thus, we suspect the discrepancies between our results and those from Knyazev {\it et al.} may come from the correction term.}
}

\cc{
Furthermore, both works of Sjostrom~\cite{15E-Sjostrom} and Witte {\it et al.}~\cite{18PP-Witte} adopted the PAW method, but their results differ, especially at temperatures higher than 2 eV.
On the one hand, Sjostrom {\it et al.} employed the PAW method with 3 valence electrons for Al, and only the $\Gamma$ point was used with a cell containing 64 Al atoms. In addition, the electron densities used in KSDFT calculations to yield Kohn-Sham orbitals were obtained from OFDFT calculations \ccc{without the self-consistent loop}, suggesting that the electron densities were approximated. \ccc{Moreover,} to compute the velocity operator, the transversal expression was adopted. 
Here, the transversal expression means directly substituting $\hat{\mathbf{v}}$ with $-i\hbar\nabla/m_e$ in Eq.~\ref{eq:KG}, considering the PAW method's ability to recover all-electron wavefunctions. 
Nonetheless, Dajdo\v{s} {\it et al.}~\cite{06B-Gajdos} pointed out that PAW typically truncates the one-center expansion, leading to inaccurate transversal expression. Consequently, they proposed incorporating a term that describes the dipole moments within the one-center sphere, known as the longitude form~\cite{06B-Gajdos}.
On the other hand, Witte {\it et al.} employed the PAW method with 11 valence electrons. The cell size and $k$-point sampling were checked to yield converged results. Importantly, the longitude form~\cite{06B-Gajdos,17PP-French} with corrections was adopted, as the different results of transversal and longitude forms were also shown by Demyanov and Knyazev's recent work {\it et al.}~\cite{22E-Demyanov}
These findings suggest that although the PAW method is generally considered an all-electron method, the truncation of partial waves introduces additional corrections that should not be overlooked.~\cite{06B-Gajdos}
}


\cc{\ccc{Recently,} Witte {\it et al.}~\cite{18PP-Witte} employed both PBE and HSE (hybrid functional) exchange-correlation functionals. 
In general, the HSE functional yields smaller electrical conductivity than PBE, but the results from the two functionals converge when the temperature exceeds 4 eV.
\ccc{Comparing with these PBE and HSE data,} our results with the non-local potential correction and NC11 agree well with the PBE data at 0.2 eV and with the HSE data at higher temperatures.
\ccc{As illustrated in Fig.~\ref{fig:cond70}(d),} unlike electrical conductivities,
\ccc{both PBE and HSE data from Witte {\it et al.}~\cite{18PP-Witte}} lead to similar thermal conductivities, and our results with \ccc{the non-local potential correction and the NC11 pseudopotential} match well with these data.
The above results highlight the importance of incorporating the non-local potential corrections when calculating the electrical and thermal conductivities of electrons.}

%


\subsection{Density of states}
\begin{figure}
	\centering
	\includegraphics[width=8.6cm]{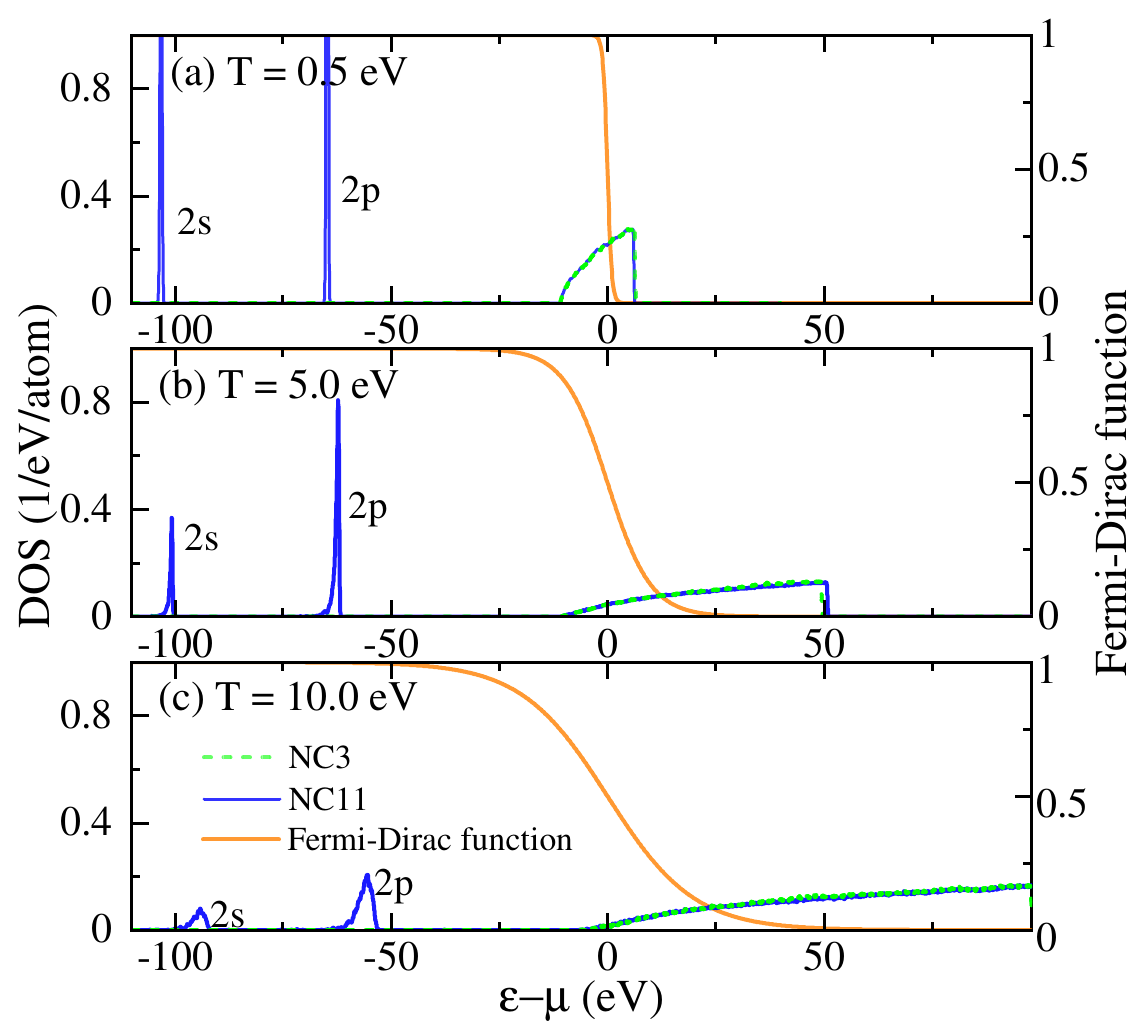}
	\caption{\cc{Density of States (DOS) of warm dense Al at a density of 2.35 $\mathrm{g/cm^3}$ and temperatures of (a) 0.5, (b) 5.0, and (c) 10.0 eV. NC11 and NC3 refer to two norm-conserving pseudopotentials with 11 and 3 valence electrons, respectively. The Fermi-Dirac function is plotted for each temperature.}}\label{fig:dos}
\end{figure}

\cc{
Fig.~\ref{fig:dos} shows the calculated density of states (DOS) of Al (2.35 $\mathrm{g/cm^3}$) at temperatures of 0.5, 5.0, and 10.0 eV, where $\mu$ is the chemical potential. \ccc{The DOS contributed by 2s and 2p electrons of Al are labeled.}
In particular, we plot the DOS of Al by using two NCPPs (NC3 and NC11). In this regard, the 2s and 2p electrons are absent \ccc{in the NC3 pseudopotential with 3 valence electrons but are included in the NC11 pseudopotential with 11 valence electrons}. Interestingly, we observe that the 2s and 2p electrons become more dispersed and shift towards the Fermi energy as temperature increases, and this physical phenomenon can only be captured by DFT calculations with the NC11 pseudopotential.
}

\cc{
Additionally, we plot the Fermi-Dirac function for each temperature. We can see that
as temperatures increase, the slope of the Fermi-Dirac function becomes smoother.
\ccc{By checking the occupations of these energy states, we find} at temperatures of 0.5 and 5 eV, the 2s and 2p orbitals of Al are fully occupied, and the electrons of the 2p orbitals start to ionize at 10 eV. \ccc{The ionization degree of the 2p orbitals is 0.36\% at the temperature of 10 eV, as evaluated from the occupation number}. This is further confirmed by seeing that the Fermi-Dirac distribution function at 10 eV begins to deviate from 1 at the peak position of the 2p orbitals.
Meanwhile, the DOS of the scattering states (3s3p and other higher-energy orbitals of Al) around the Fermi level shifts to higher energies as temperature arises, which is captured by both NC3 and NC11 pseudopotentials. In conclusion, even though core electrons are almost non-ionized at the temperature of 5 eV, \ccc{the energy levels of 2s and 2p electrons move towards higher energies and affect the computed conductivities, suggesting that} a pseudopotential with 11 valence electrons is {a more appropriate choice than the NC3 pseudopotential} for warm dense Al.
}

\begin{figure*}
	\centering
	\includegraphics[width=18cm]{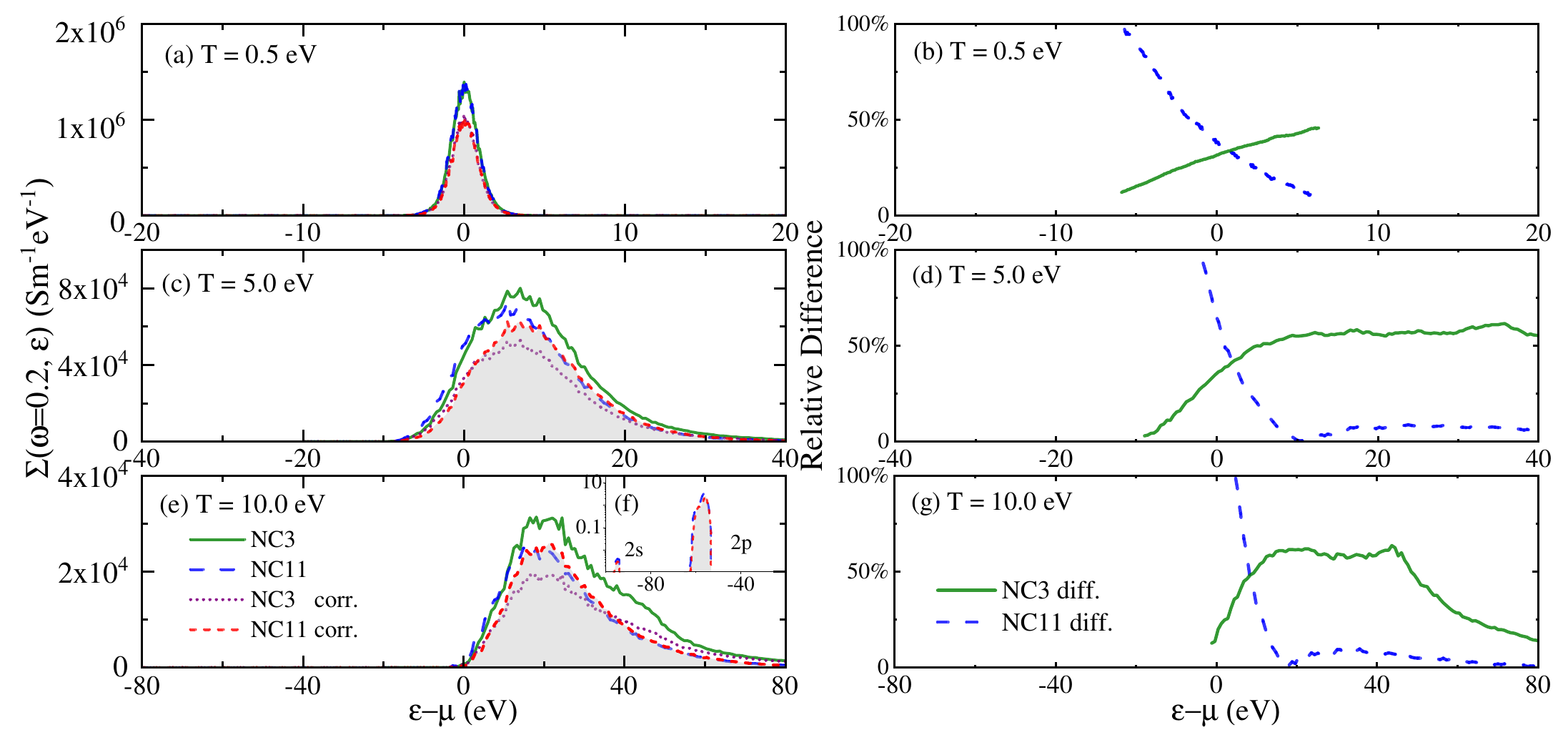}
	\caption{Decomposed electrical conductivity $\Sigma(\omega,\varepsilon)$ (as defined in Eq.~\ref{DEC}) of warm dense Al with respect to different energies $\varepsilon$ at temperatures ($T$) of (a) 0.5, (c) 5.0, and (d) 10.0 eV. Here $\mu$ is the chemical potential. The inset (f) shows the decomposed electrical conductivity contributed by the 2s and 2p orbitals of Al.
		Besides, we plot the relative difference of $\Sigma(\omega, \varepsilon)$ between results with and without the non-local potential correction for temperatures of (b) 0.5, (d) 5.0, and (g) 10.0 eV.
		The density of Al is 2.35 $\mathrm{g/cm^3}$ and the frequency $\omega$ is chosen to be 0.2 eV.
		The area under each curve represents the zero-frequency electrical conductivity.
		NC11 and NC3 refer to two norm-conserving pseudopotentials with 11 and 3 valence electrons, respectively. The term ``corr." depicts the non-local potential correction.
		``NCx diff.“ represents the relative difference computed by the formula of $|\Sigma_\mathrm{NCx} - \Sigma_\mathrm{NCx\ corr.}|/\Sigma_\mathrm{NCx\ corr.}$, where x is 3 or 11.
 }\label{fig:contribute}
\end{figure*}

\subsection{Decomposed electrical conductivity}

\cc{
We define a new quantity named decomposed electrical conductivity $\Sigma(\omega,\varepsilon)$ to analyze the contributions of different orbitals to the electrical conductivity. The formula of $\Sigma$ is
\begin{equation}
	\begin{aligned}
&\Sigma(\omega,\varepsilon)=\frac{2\pi e^2}{3\omega\Omega}\sum_{ij\alpha\mathbf{k}}W(\mathbf{k})
	|\langle\Psi_{i\mathbf{k}}|\hat{v}_\alpha|\Psi_{j\mathbf{k}}\rangle|^2\\
	&\times[f(\epsilon_{i\mathbf{k}})-f(\epsilon_{j\mathbf{k}})]\delta(\epsilon_{j\mathbf{k}}-\epsilon_{i\mathbf{k}}-\hbar\omega)\delta(\epsilon_{i\mathbf{k}}-\varepsilon),
	\end{aligned}
 \label{DEC}
\end{equation}
which satisfies the relationship of
\begin{equation}
	\sigma(\omega) = \int \Sigma(\omega,\varepsilon) \mathrm{d}\varepsilon.  
\end{equation}
Here $\sigma(\omega)$ is the frequency-dependent electrical conductivity.
The results of $\Sigma(\omega,\varepsilon)$ as computed by two different pseudopotentials (NC3 and NC11) with and without the non-local potential corrections are displayed in Fig.~\ref{fig:contribute}, where $\mu$ is the chemical potential.
Since conductivities are calculated by extrapolating dynamic conductivities at small frequencies, we choose $\omega$=0.2 eV as a representative frequency to investigate the effects of pseudopotentials and non-local potential corrections. According to the above formulas, the area of each line in Fig.~\ref{fig:contribute} represents the value of $\sigma(\omega)$.
}
As shown in Fig.~\ref{fig:cond35}, the consideration of \ccc{the non-local corrections and pseudopotentials with different numbers of electrons} results in different behaviors at low and high temperatures. Here, we calculate $\Sigma(\omega,\varepsilon)$ at a temperature of 0.5 eV to represent low temperatures, and at temperatures of 5.0 and 10.0 eV to represent high temperatures. As depicted in Fig.~\ref{fig:dos}, the 2p electrons \ccc{are not ionized at 5.0 eV but begin to ionize at 10.0 eV.} We have the following findings.

On the one hand, at the low temperature of 0.5 eV, as shown in Fig.~\ref{fig:contribute}(a), we observe that the decomposed electrical conductivity has peaks around the Fermi surface, indicating that only electrons around the Fermi surface ($\varepsilon-\mu$=0) contribute to the electrical conductivity. 
This can be understood by the term of $f(\epsilon_{i\mathbf{k}})-f(\epsilon_{j\mathbf{k}})$ included in the formula of $\Sigma(\omega,\varepsilon)$, and $\partial f(\varepsilon)/\partial \varepsilon$ has the largest value at the peak.
We also find the distribution of $\Sigma(\omega,\varepsilon)$ using the NC11 and NC3 pseudopotentials is similar, indicating that the 2s and 2p electrons in pseudopotentials \ccc{barely} affect the electrical conductivity of Al at 0.5 eV. Meanwhile, the inclusion of non-local potential correction significantly reduces the conductivity, highlighting the importance of considering this correction.

Furthermore, Fig.~\ref{fig:contribute}(b) illustrates the relative differences between $\Sigma$ with and without the non-local potential correction at 0.5 eV. \ccc{We observe} the relative difference of the NC3 pseudopotential increases with the rise of eigenenergies of electrons, while that of NC11 decreases. This implies the correction becomes more prominent for NC3 than NC11 as shown in Fig.~\ref{fig:cond35}(b) because more electrons \ccc{are} excited to high energies with the rise of the temperature.

On the other hand, as the temperature \ccc{respectively} rises to 5 eV and 10 eV in Fig.~\ref{fig:contribute}(b) and (e), the peaks of $\Sigma$ shift towards higher energies because the velocity matrix term becomes substantially larger at high-energy bands.

\ccc{However,} the electrical conductivities obtained from the NC3 and NC11 pseudopotentials exhibit substantial discrepancies when the energy $\varepsilon-\mu$ is around the peak of distributions. 
While the density of states (DOS) of Al calculated by NC3 and NC11 are similar in Fig.~\ref{fig:dos}, the variations in the velocity matrix values calculated with corresponding states, whether with or without non-local potential corrections, contribute to these discrepancies.
This suggests that the scattering states of the two systems are inherently different.
Consequently, even though the 2p electrons of Al are either still occupied at 5 eV or barely ionized at 10 eV, the total conductivities show significant differences at these higher temperatures.
Panel (f) further confirms that the \ccc{contributions from 2s and 2p electrons are} much smaller compared to \ccc{the} scattering states.
These indicate that including the 2s and 2p orbitals in the NCPP is critical in accurately calculating correct scattering states and predicting the conductivities.

Besides, the relative difference of the NC11 pseudopotential continues to decrease to a small value due to the predominance of the kinetic energy, indicating that the correction becomes less significant with more excited electrons and a shift of peak to higher energies as the temperature rises.
In contrast, for the NC3 pseudopotential, \ccc{the relative differences are in general larger than those from the NC11 pseudopotential, except when $\varepsilon-\mu$ is small.}



\subsection{The Lorenz number}
\begin{figure}
	\centering
	\includegraphics[width=8.6cm]{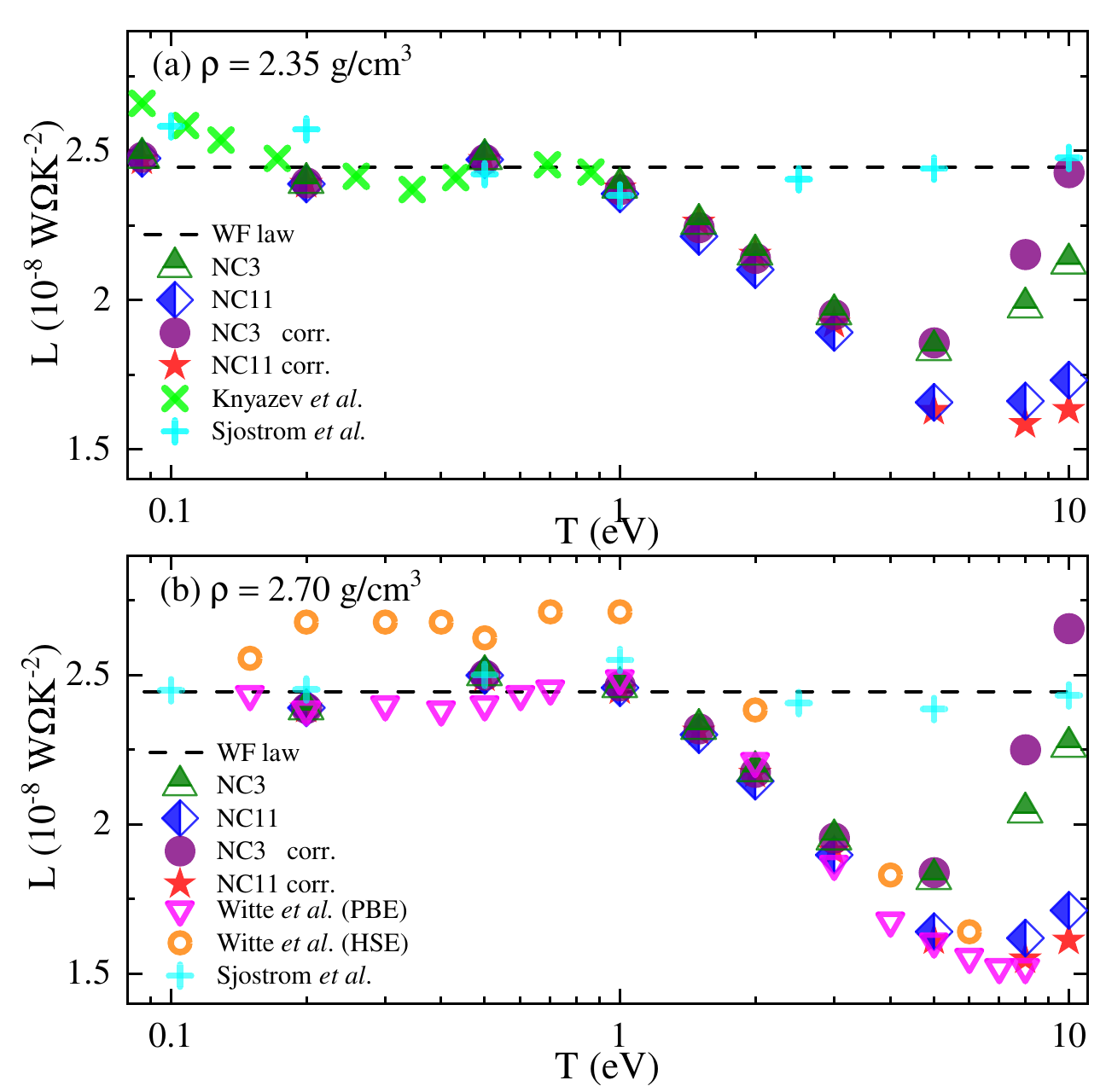}
	\caption{\cc{The Lorenz numbers of warm dense aluminum at temperatures ranging from 0.086 to 10 eV and densities of (a) 2.35 $\mathrm{g/cm^3}$ and (b) 2.70 $\mathrm{g/cm^3}$. NC11 and NC3 are two norm-conserving pseudopotentials. The label corr. means the non-local potential correction is included. The cell contains 256 Al atoms for temperatures no more than 5 eV and 108 atoms for higher temperatures.} Besides, \ccc{first-principles data} from Knyazev {\it et al.}~\cite{13CMS-Knyazev}, Sjostrom {\it et al.}~\cite{15E-Sjostrom}, and Witte {\it et al.}~\cite{18PP-Witte} are also presented for comparison.}
 \label{fig:Lorenz}
\end{figure}

\cc{
The Lorenz number ~\cite{53AP-Franz,72AP-Lorenz} characterizes the relationship between the electrical and thermal conductivities and is still not well understood in the warm dense matter region.
It is defined as
\begin{equation}
	L=\frac{\kappa}{\sigma T},
\end{equation}
where $T$ is the temperature. Here $\kappa$ and $\sigma$ refer to the thermal conductivity and the DC electrical conductivity, respectively. According to the Wiedemann-Franz law~\cite{53AP-Franz,72AP-Lorenz,01-Ziman}, $L$ is a constant for highly degenerate electrons, approximately equal to 
$2.445\times10^{-8}~\mathrm{W\Omega K^{-2}}$.
According to our calculation results shown in Fig.~\ref{fig:Lorenz}, we have the following findings.
}

\cc{
First, at temperatures below 3 eV, the Lorenz number calculated with the NC3 and NC11 pseudopotentials exhibit the same trend. However, as the temperature further increases, the results differ due to the invalidity of the frozen-core approximation used in the NC3 pseudopotential.
}

\cc{
Second, the Lorents number obtained by the NC11 pseudopotential with and without the non-local potential corrections are close, indicating that the corrections influence the electrical and thermal conductivities in the same proportion.
In contrast, by utilizing the NC3 pseudopotential at the temperature of 8 eV and higher, the non-local potential corrections become pronounced.
}

\cc{
Third, at temperatures up to 1 eV, the Lorenz number agrees well with the Wiedemann-Franz (WF) law, indicating the inherent relationship between electrical conductivity and thermal conductivity is valid. Meanwhile, the degeneracy parameter, defined as the ratio of temperature to the Fermi temperature $T/T_F$, is 0.086 at the density of 2.7 $\mathrm{g/cm^3}$ and 0.094 at 2.35 $\mathrm{g/cm^3}$, indicating a high degree of degeneracy in these Al systems.
}

\cc{
Fourth, importantly, as the temperature exceeds 1 eV, which falls into the intermediate degenerate range, the Lorenz number starts to decrease and reaches a minimum at the temperature of approximately 8 eV. 
Specifically, at the density of 2.7g/cm$^3$, the Lorenz number is 36.7\% lower than that predicted by the WF law, while at the density of 2.35 g/cm$^3$, it is 35.1\% lower.
The observed trend of the Lorentz number demonstrates that in this region, we cannot calculate thermal conductivity directly from the electrical conductivity using the Wiedemann-Franz law.
}

Fifth, we compared our results with those of previous works~\cite{05B-Recoules,13CMS-Knyazev,14PP-Knyazev,15E-Sjostrom,18PP-Witte}. 
The results obtained by Recoules {\it et al.}~\cite{05B-Recoules} \ccc{followed the Wiedemann-Franz law} in the temperature range of 70 to 10000 K (0.006 to 0.862 eV) at a density of 2.35 g/cm$^3$ (not shown in Fig.~\ref{fig:Lorenz}),
\ccc{agreeing well with our results.}
The PBE results of Witte {\it et al.}~\cite{18PP-Witte} also match well with our results when using the NC11 pseudopotential with non-local potential corrections. However, the HSE results predict a higher Lorenz number. It should be noted that all of these results \ccc{have taken} into account \ccc{the} velocity matrix corrections. 
\ccc{Next,} Knyazev {\it et al.}~\cite{13CMS-Knyazev} \ccc{reported} slightly higher \ccc{values of the Lorents number} at low temperatures. 
In addition, we also calculate the Lorenz number using $\sigma$ and $\kappa$ values from Sjostrom {\it et al.}~\cite{15E-Sjostrom}. Their results agree well with the Wiedemann-Franz law for all temperatures below 10 eV, yielding unfavorable results as it fails to exhibit a decrease in the Lorenz number at higher temperatures as expected. \ccc{This may be attributed to the approximations used in their work, such as the non-self-consistent electronic density obtained} by OFDFT.

\section{Conclusions}
\cc{
Accurate prediction of electron transport coefficients is vital for comprehending warm dense matter. The Kubo-Greenwood formula is extensively employed to calculate the electrical and thermal conductivities of electrons. Nonetheless, the existence of non-local terms in the pseudopotentials and the use of the frozen-core approximation in first-principles calculations lead to inaccuracy when the Kubo-Greenwood formula is adopted.}

\cc{
In this work, we implemented the non-local potential corrections with the use of norm-conserving pseudopotentials and plane wave basis under periodic boundary conditions within the ABACUS package. We calculated both electrical and thermal conductivities of liquid Al at 1000 K and warm dense Al with temperatures ranging from 0.2 to 10 eV. To ensure converged results, we used converged numbers of atoms and $k$-points in our calculation.}

\cc{We examined the electrical and thermal conductivities of liquid Al at a temperature of 1000 K. By incorporating the non-local potential corrections, we obtained results that agreed well with the experimental data. 
Next, we simulated warm dense Al with temperatures ranging from 0.2 eV to 10.0 eV at a density of 2.35 $\mathrm{g/cm^3}$.
Compared to the results without using the non-local potential corrections, we observed lower conductivities when the corrections were applied. 
Notably, two different norm-conserving pseudopotentials, NC3 and NC11, yielded similar electrical and thermal conductivities at low temperatures. However, as temperatures exceeded 2 eV, discrepancies emerged due to the inadequate frozen core approximation of the 2s and 2p electrons of Al.
We also simulated Al at a density of 2.70 $\mathrm{g/cm^3}$.
The results exhibited similar behaviors to those at 2.35 $\mathrm{g/cm^3}$ and showed reasonable agreement with available experimental data. We also compared our calculated results to those obtained by previous works.
}

\cc{
To gain further insights into the computed conductivities, we analyzed the DOS and the decomposed electrical conductivity at temperatures of 0.5, 5, and 10 eV. We found the core electrons are almost non-ionized at the temperature of 5 eV. However, the core electrons affect the high-energy electronic states, resulting in substantially different conductivities as compared to the N3 pseudopotential.
In conclusion, the NC11 pseudopotential provides a more accurate prediction than the NC3 pseudopotential for warm dense Al.}
%
%

%
\cc{
In summary, our findings emphasize the importance of incorporating non-local potential corrections when calculating the conductivities of electrons using pseudopotentials. Additionally, the choice of pseudopotential is crucial in accurately determining conductivities, and the frozen-core approximation becomes invalid for calculating conductivities at temperatures before the ionization of corresponding core electrons.}
However, \ccc{limited by high computational costs for electrons at extremely high temperatures,} traditional KSDFT only allows conductivity calculations for warm dense Al up to a few tens of eV.
To overcome this limitation, recent advancements in stochastic DFT~\cite{13L-Baer,18B-Cytter} and mixed stochastic-deterministic DFT~\cite{20L-White,22B-Liu,22JPCM-White} present promising approaches for addressing this challenge.

\acknowledgements
This work is supported by the National Science Foundation of China under Grant No.12122401 and No.12074007. We thank the Electronic Structure Team at AI for Science Institute in Beijing for contributing to the ABACUS package. We thank Tao Chen from Peking University for reviewing this manuscript.
Part of the numerical simulations was performed on the High-Performance Computing Platform of CAPT.

\bibliography{nonlocal}

\end{document}